\documentstyle[preprint,aps]{revtex}

\begin{document}
\draft
\title{Implementation of non-local operations for arbitrary high-dimensional
systems with qubit quantum channel}
\author{Hao-Sheng Zeng\thanks{%
E-mail adress: hszeng@hunnu.edu.cn}, Yong-Guang Shan, Jian-Jun Nie and
Le-Man Kuang}
\address{Department of Physics, Hunan Normal University, Changsha\\
410081, People's Republic of China}
\maketitle
\address{}

\begin{abstract}
We propose a method to implement a kind of non-local operations between
spatially separated two systems with arbitrary high-dimensions by using only
low-dimensional qubit quantum channels and classical bit communications. The
result may be generalized straightforwardly to apply for multiple systems,
each of them with arbitrary dimensions. Compared with existed approaches,
our method can economize classical resources and the needed low-dimensional
quantum channels may be more easily established in practice. We also show
the construction of the non-local quantum XOR gate for qutrit systems in
terms of the obtained non-local operations as well as some single qutrit
local gates.
\end{abstract}

\pacs{PACS number(s): 03.67.Mn, 03.65.Ta, 03.65.Ud}

\vskip 1cm

\narrowtext

Entanglement is a very mystical and important phenomenon in quantum physics
which has been studied extensively and deeply in theoretics and experiments.
It has been playing a pivotal role in many aspects of quantum information
and quantum computation, such as teleportation, dense coding , quantum
cryptography and distributed quantum computation \cite{Nielsen}. Recently,
an interesting application of entanglement has been put forward which
involves the implementation of non-local quantum gates or interactions \cite
{Eisert,Kraus,Collins,Huelga,Reznik,Cirac,Dura,Durb,Huang,Berry,Chen}
between spatially separated systems, by using a given resource of entangled
states and applying local operations as well as classical communication.
This subject is essentially important to realize the distributed quantum
computation\cite{Cirac1}, because all quantum unitary operations between
distributed systems can be realized by combining this non-local gate with
local operations. The important aspect to implement the non-local gates is
to increase the efficiency\cite{Cirac,Berry,Chen}, given the limited
resources including entangled qubits and classical bits, which is still
worthwhile to be studied.

In addition, quantum entanglement in a high-dimensional Hilbert space have
potential applications in quantum information and quantum computation\cite
{Stephen,Thew}. Since particles in high-dimensional systems, i.e., qudits,
can carry more information than qubits, it can increase information flux in
quantum communication. It can also increase security \cite{Bruss} in quantum
cryptography by exploiting the high-dimensional systems. Further,
high-dimensional systems have advantages in fault-tolerant quantum
computation and quantum error-correction\cite{Knill}. Therefore, extending
non-local operations to high-dimensional systems is also important.
Reference \cite{Reznik} presented a method to implement a kind of non-local
operations between multiple systems with arbitrary discrete dimensions by
employing state-operator tool. However, it requires to consume
high-dimensional entanglement resources, as well as employ classical-dit
communication. In this paper, we show an approach for constructing a kind of
non-local unitary operations between two spatially separated multi-level
systems, by using only one low-dimensional quantum channel together with
2-cbit communication. Moreover, it can also be generalized straightforwardly
to apply for multiple systems, each of them with arbitrary (maybe different)
dimensions. The method has the advantages of economizing classical resources
and the needed low-dimensional quantum channels maybe more easily
established in practice. In addition, we also discuss the applications of
this kind of non-local operations in producing high-dimensional non-local
entanglement states and performing distributed quantum computation.

Let us consider two-particle systems $A$ and $B$ which are remotely
separated in space, the partners Alice and Bob previously share an
accessorial entangled state 
\begin{equation}
\left| \Psi _{a{b}{b}_{1}}\right\rangle =\lambda _{0}\left|
0_{a}0_{b}0_{b_{1}}\right\rangle +\lambda _{1}\left|
0_{a}0_{b}1_{b_{1}}\right\rangle +\lambda _{2}\left|
1_{a}1_{b}0_{b_{1}}\right\rangle +\lambda _{3}\left|
1_{a}1_{b}1_{b_{1}}\right\rangle .
\end{equation}
Here, the accessorial particle $a$ belongs to Alice and $b$, $b_{1}$ to Bob.
For simplicity, we assume that the coefficient's $\lambda _{i}$ are
non-negative real numbers and normalized as: $\sum_{i=0}^{3}|\lambda
_{i}|^{2}=1$. We can easily get the entanglement of this state as 
\begin{equation}
E(\left| \Psi _{a{b}{b_{1}}}\right\rangle )=-H\log H-(1-H)\log (1-H).
\end{equation}
Where $H=\lambda _{0}^{2}+\lambda _{1}^{2}$ and the base of logarithm $\log
(x)$ is taken to be 2.

In order to obtain the expected non-local operation on particles $A$ and $B$%
, Alice and Bob firstly perform the following local unitary operations
respectively [A kind of control-$U$ operations with $a$, $b_{1}$ the control
bits and $A$, $B$ the target dits respectively.] 
\begin{eqnarray}
U_{aA} &=&\left| 0_{a}\right\rangle \left\langle 0_{a}\right| \otimes {I_{A}}%
+i\left| 1_{a}\right\rangle \left\langle 1_{a}\right| \otimes {U_{A}}, 
\nonumber \\
U_{{b_{1}}B} &=&\left| 0_{b_{1}}\right\rangle \left\langle 0_{b_{1}}\right|
\otimes {I_{B}}+\left| 1_{b_{1}}\right\rangle \left\langle 1_{b_{1}}\right|
\otimes {U_{B}}.
\end{eqnarray}
Here $U_{aA}$ ($U_{{b_{1}}B}$) denotes that the unitary operation is
performed between accessorial particle $a$ ($b_{1}$) and target particle{\em %
\ }$A$ ($B$). After these local unitary operations, they get a stator of the
following form 
\begin{equation}
S_{1}=\lambda _{0}\left| 0_{a}0_{b}0_{b_{1}}\right\rangle +\lambda
_{1}\left| 0_{a}0_{b}1_{b_{1}}\right\rangle \otimes {U_{B}}+i\lambda
_{2}\left| 1_{a}1_{b}0_{b_{1}}\right\rangle \otimes {U_{A}}+i\lambda
_{3}\left| 1_{a}1_{b}1_{b_{1}}\right\rangle \otimes {U_{A}U_{B}}.
\end{equation}
Then Alice performs a measurement of $\sigma _{xa}$ on stator $S_{1}$ with
respect to the accessorial particle $a$, and transmits his measurement
outcome (through one classical bit) to Bob. Following this message, Bob will
do nothing or perform an operation of $\sigma _{zb}$ on his accessorial
particle {\em $b$}, conditional on his received information $1$ or $-1$.
After this process, the stator becomes

\begin{equation}
S_{2}=\lambda _{0}\left| 0_{b}0_{b_{1}}\right\rangle +\lambda _{1}\left|
0_{b}1_{b_{1}}\right\rangle \otimes {U_{B}}+i\lambda _{2}\left|
1_{b}0_{b_{1}}\right\rangle \otimes {U_{A}}+i\lambda _{3}\left|
1_{b}1_{b_{1}}\right\rangle \otimes {U_{A}U_{B}}
\end{equation}
In order to eliminate accessorial particles {\em $b$} and {\em $b_{1}$ }and
get the expected non-local operation, Bob now collectively measures his
particles {\em $b$} and {\em $b_{1}$} in the following Bell-like basis 
\begin{eqnarray}
|B_{1}\rangle &=&\cos \alpha |0_{b}0_{b_{1}}\rangle +\sin \alpha
|1_{b}1_{b_{1}}\rangle ,  \nonumber \\
|B_{2}\rangle &=&\cos \alpha |1_{b}1_{b_{1}}\rangle -\sin \alpha
|0_{b}0_{b_{1}}\rangle ,  \nonumber \\
|B_{3}\rangle &=&\cos \beta |0_{b}1_{b_{1}}\rangle +\sin \beta
|1_{b}0_{b_{1}}\rangle ,  \nonumber \\
|B_{4}\rangle &=&\cos \beta |1_{b}0_{b_{1}}\rangle -\sin \beta
|0_{b}1_{b_{1}}\rangle ,
\end{eqnarray}
where $\alpha $ and $\beta $ are real numbers. The corresponding probability
that $S_{2}$ projected onto each basis is

\begin{eqnarray}
P_{1} &=&\lambda _{0}^{2}\cos ^{2}\alpha +\lambda _{3}^{2}\sin ^{2}\alpha , 
\nonumber \\
P_{2} &=&\lambda _{0}^{2}\sin ^{2}\alpha +\lambda _{3}^{2}\cos ^{2}\alpha , 
\nonumber \\
P_{3} &=&\lambda _{1}^{2}\cos ^{2}\beta +\lambda _{2}^{2}\sin ^{2}\beta , 
\nonumber \\
P_{4} &=&\lambda _{1}^{2}\sin ^{2}\beta +\lambda _{2}^{2}\cos ^{2}\beta ,
\end{eqnarray}
respectively. After getting one of the resulting operators, Bob sends 1-cbit
information to inform Alice whether she will perform $U_{A}$ (for outcomes $%
P_{2}$ and $P_{4})$ or not (for outcomes $P_{1}$ and $P_{3}$). At the same
time, Bob will also perform a local operation of $U_{B}$ (for outcomes $%
P_{2} $ and $P_{3}$) or do nothing (for outcomes $P_{1}$ and $P_{4}$). Then
they can obtain respectively non-local unitary operators 
\begin{eqnarray}
U_{1AB} &=&P_{1}^{-1/2}[\lambda _{0}\cos \alpha +i\lambda _{3}\sin \alpha {%
U_{A}}U_{B}],  \nonumber \\
U_{2AB} &=&P_{2}^{-1/2}[\lambda _{3}\cos \alpha +i\lambda _{0}\sin \alpha {%
U_{A}}U_{B}],  \nonumber \\
U_{3AB} &=&P_{3}^{-1/2}[\lambda _{1}\cos \beta +i\lambda _{2}\sin \beta {%
U_{A}}U_{B}],  \nonumber \\
U_{4AB} &=&P_{4}^{-1/2}[\lambda _{2}\cos \beta +i\lambda _{1}\sin \beta {%
U_{A}}U_{B}].
\end{eqnarray}
If we choose $\lambda _{0}=\lambda _{3}$, $\lambda _{1}=\lambda _{2}$
[Eq.(2) tells us that in this case, the entanglement of the quantum channel
of eq.(1) is maximal.] and $\alpha =\beta =\xi $, then we get $%
U_{1AB}=U_{2AB}=U_{3AB}=U_{4AB}$ which have the following general form 
\begin{equation}
U_{AB}(\xi )=\exp {[i\xi {U_{A}}U_{B}]}.\quad \xi \in [0,2\pi ].
\end{equation}
Thus, we obtain a non-local unitary operation between spatially separated
systems $A$ and $B$. The probability to produce this non-local operation is
apparently one, and the angle $\xi $ can be adjusted at will by Bob alone.
Note that the local operators $U_{A}$ and $U_{B}$ for particles $A$ and $B$
are not only unitary, but also Hermitian.

It is worthwhile to point out that our approach to produce high-dimensional
non-local operation can be regarded as a straightforward generalization of
the method to produce non-local qubit operation \cite{Chen}. Especially, for
qubit systems, we can choose $U_{A}=\sigma _{nA}$ and $U_{B}=\sigma _{nB}$
with $n=x,y,z$, then eq.(9) reduces to the well known form of $\exp (i\xi
\sigma _{nA}\sigma _{nB})$, which represents a general non-local operation
in the sense that, along with single qubit local operations, it can realize
any expected unitary operation between spatially separated qubits $A$ and $B$%
.

Note also that, like the results presented by many authors \cite{Reznik,Chen}%
, one of the prerequisites to determinately produce the non-local operation
of eq.(9) is the utilizing of a maximally entangled quantum channel.
However, imitating the processing offered by reference \cite{Chen} and at
the expense of successful probability, we can also get this non-local
operation through the use of a lower entangled quantum channel.

Our result has extensive applications. It can be applied to the case of any
high-dimensional systems, including $A$ and $B$ with different dimensions.
In our method, we employ only low-dimensional qubit-entanglement resource.
Thus compared with the approach\cite{Reznik} that using high-dimensional
entanglement resources to produce the corresponding non-local operations, it
is more simple and the needed entanglement resources are more easily
obtained. Moreover, the classical consumptions in our scheme are always two
cbits (i.e., bidirectional bit communication between Alice and Bob), rather
than two classical dits\cite{Reznik}. Thus, it will economize many classical
resources when higher dimensional systems involved.

The non-local operation of eq.(9) give us an appropriate room to realize a
variety of interactions. In order to obtain a concrete non-local operation
between two $d$-dimensional systems, it is needed for finding an adequate
unitary as well as Hermitian operator $U_{d}$ for systems $A$ and $B$. If we
regard the basis $\{\left| s\right\rangle ;~s=0,1,\ldots ,d-1\}$ of a $d$%
-dimensional system as the eigenstates of angular momentum operator $J_{z}$
of a spin system with eigenvalues $-j,\ldots ,j$ respectively ( $j$ is the
highest quantum number which is an integer or half-integer and satisfies $%
d=2j+1$), then a simple and intuitionistic choice for unitary and Hermitian
operator $U_{d}$ is

\begin{equation}
U_{d}=\left\{ 
\begin{array}{l}
\exp (i\pi J_{n}),\qquad \ ~d\quad \text{odd} \\ 
i\exp (i\pi J_{n}),\qquad d\quad \text{even}
\end{array}
.\right.
\end{equation}
\noindent

Of course, in a $d$-dimensional system, the choice for $U_{d}$ is not single
in general. There may exist many other forms of choices. As an example, we
write out the general form of $U_{3}$ for a three-dimensional system, i.e.,
qutrit, in the basis $\{\left| 0\right\rangle ,\left| 1\right\rangle ,\left|
2\right\rangle \}$

\begin{equation}
U_{3}=\left[ 
\begin{array}{lll}
a_{1} & b_{1}e^{i\varphi _{1}} & b_{2}e^{i\varphi _{2}} \\ 
b_{1}e^{-i\varphi _{1}} & a_{2} & b_{3}e^{i\varphi _{3}} \\ 
b_{2}e^{-i\varphi _{2}} & b_{3}e^{-i\varphi _{3}} & a_{3}
\end{array}
\right]
\end{equation}
with parameters $a_{i}$, $b_{i}$ and phases $\varphi _{i}$ ($i=1,2,3$) are
all real. Through the investigation of the unitarity of $U_{3}$, we find
that the phases $\varphi _{i}$ only need satisfying one of the two conditions

\begin{equation}
\varphi _{1}=\varphi _{2}-\varphi _{3},
\end{equation}

\begin{equation}
\varphi _{1}=\varphi _{2}-\varphi _{3}+\pi ,
\end{equation}
and the parameters $a_{i}$, $b_{i}$ should satisfy the following set of
equations

\begin{equation}
\left\{ 
\begin{array}{l}
a_{1}^{2}+b_{1}^{2}+b_{2}^{2}=1, \\ 
a_{2}^{2}+b_{1}^{2}+b_{3}^{2}=1, \\ 
a_{3}^{2}+b_{2}^{2}+b_{3}^{2}=1, \\ 
(a_{1}+a_{2})b_{1}\pm b_{2}b_{3}=0, \\ 
(a_{1}+a_{3})b_{2}\pm b_{1}b_{2}=0,
\end{array}
\right.
\end{equation}
where the `+' in front of $b_{2}b_{3}$ and $b_{1}b_{2}$ corresponds to
eq.(12), and `-' corresponds to eq.(13). Further investigation suggests that
this set of equations has, and nay has, multiple sets of real number
solutions. [Note that the number of variables are larger than that of
equations.]

The non-local operation of eq.(9) presents a means of producing
higher-dimensional non-local entanglement states in terms of
lower-dimensional qubit-entanglement resources. As an example, let us show
how to produce the 3-dimensional non-local maximally entangled state

\begin{equation}
{\textstyle {1 \over \sqrt{3}}}%
[\left| 0_{A}0_{B}\right\rangle +\left| 1_{A}1_{B}\right\rangle +\left|
2_{A}2_{B}\right\rangle ].
\end{equation}
The process for the creation of this state requires the twice uses of
eq.(9). Assume that the initial state of the bipartite system is $%
|0_{A}0_{B}\rangle $. First, we choose $U_{i}=\left| 1_{i}\right\rangle
\left\langle 0_{i}\right| +\left| 0_{i}\right\rangle \left\langle
1_{i}\right| +\left| 2_{i}\right\rangle \left\langle 2_{i}\right| $ with $%
i=A $ or $B$, and $\xi =\arcsin (\sqrt{2/3})$, then the action of eq.(9) on
this initial state produces 
\begin{equation}
\sqrt{%
{\textstyle {1 \over 3}}%
}\left| 0_{A}0_{B}\right\rangle +i\sqrt{%
{\textstyle {2 \over 3}}%
}\left| 1_{A}1_{B}\right\rangle .
\end{equation}
Afterward, we let $U_{A}=\left| 0_{A}\right\rangle \left\langle 0_{A}\right|
+\left| 1_{A}\right\rangle \left\langle 2_{A}\right| +\left|
2_{A}\right\rangle \left\langle 1_{A}\right| $, $U_{B}=\left|
0_{B}\right\rangle \left\langle 0_{B}\right| +i\left| 1_{B}\right\rangle
\left\langle 2_{B}\right| -i\left| 2_{B}\right\rangle \left\langle
1_{B}\right| $ and $\xi =\pi /4$, it then produces

\begin{equation}
{\textstyle {1 \over \sqrt{3}}}%
[e^{i\pi /4}\left| 0_{A}0_{B}\right\rangle +i\left| 1_{A}1_{B}\right\rangle
+i\left| 2_{A}2_{B}\right\rangle ].
\end{equation}

Finally, an action of local operation of $e^{i\pi /4}\left|
0_{A}\right\rangle \left\langle 0_{A}\right| +\left| 1_{A}\right\rangle
\left\langle 1_{A}\right| +\left| 2_{A}\right\rangle \left\langle
2_{A}\right| $ will enable us to get the desired state of eq.(15). Note that
the creation of this non-local 3-dimensional maximally entangled state
involves the twice uses of eq.(9), which indicates the consumption of two
maximally entangled qubit-resource states. This is consistent with the fact
that local unitary operations and classical communication can not increase
the entanglement \cite{Vedral}. Since the Von Neumann entropy is $\log 3$
for the state of eq.(15), and $1$ for the entanglement resource used above,
thus single utilizing of eq.(9) is not likely to produce the state of
eq.(15).

The non-local operation of eq.(9) may also play an important role in
multi-level distributed quantum computation and non-local quantum
information processing. As an example, let us show that how to construct a
XOR gate in terms of non-local operation of eq.(9), along with single
particle local operations. For simplicity, our discussion will be only
limited to the case of 3-dimensional systems, i.e., qutrits. The XOR gate
between qutrits $A$ and $B$ in this case is defined as

\begin{equation}
XOR_{AB}\left| j\right\rangle _{A}\left| k\right\rangle _{B}=\left|
j\right\rangle _{A}\left| j\oplus k\right\rangle _{B},
\end{equation}
where the ``$\oplus $'' operation now indicates addition modulo 3. It has
been pointed out \cite{Klimov} that the XOR operation can be decomposed into
three operations

\begin{equation}
XOR_{AB}=F_{A}P_{AB}F_{A}^{-1}
\end{equation}
where the Fourier transform for one qutrit is defined as

\begin{equation}
F\left| j\right\rangle =\frac{1}{\sqrt{3}}\sum_{l=0}^{2}e^{i2\pi jl/3}\left|
l\right\rangle ,~j=0,1,2,
\end{equation}
and the phase gate $P_{AB}$ between qutrits $A$ and $B$ as

\begin{equation}
P_{AB}\left| j\right\rangle _{A}\left| k\right\rangle _{B}=\exp (i2\pi
jk/3)\left| j\right\rangle _{A}\left| k\right\rangle _{B}.
\end{equation}
Usually, we assume that the local unitary operations are easily obtained.
Thus the task of finding a non-local XOR$_{AB}$ gate is equivalent to that
of finding a non-local phase operation $P_{AB}$ which, we will show below,
can be constructed by the use of non-local operation of eq.(9) four times,
along with some single qutrit operations. For clarity, we rewrite $P_{AB}$ as

\begin{eqnarray}
P_{AB} &=&\left| 00\right\rangle \left\langle 00\right| +\left|
01\right\rangle \left\langle 01\right| +\left| 02\right\rangle \left\langle
02\right| +\left| 10\right\rangle \left\langle 10\right| +e^{i2\pi /3}\left|
11\right\rangle \left\langle 11\right|  \nonumber \\
&&+e^{i4\pi /3}\left| 12\right\rangle \left\langle 12\right| +\left|
20\right\rangle \left\langle 20\right| +e^{i4\pi /3}\left| 21\right\rangle
\left\langle 21\right| +e^{i2\pi /3}\left| 22\right\rangle \left\langle
22\right| .
\end{eqnarray}
Where we introduce abbreviation $\left| ij\right\rangle \equiv \left|
i\right\rangle _{A}\left| j\right\rangle _{B}$ with $i,j=0,1,2$ to denote
the basis of combined system. Now we first put $U_{A}=\left|
0_{A}\right\rangle \left\langle 0_{A}\right| +\left| 1_{A}\right\rangle
\left\langle 1_{A}\right| -\left| 2_{A}\right\rangle \left\langle
2_{A}\right| $, $U_{B}=\left| 0_{B}\right\rangle \left\langle 0_{B}\right|
-\left| 1_{B}\right\rangle \left\langle 1_{B}\right| +\left|
2_{B}\right\rangle \left\langle 2_{B}\right| $ and $\xi =\gamma $, then the
non-local operation of eq.(9) becomes

\begin{eqnarray}
U_{AB}^{\prime }(\gamma ) &=&e^{i\gamma }\left| 00\right\rangle \left\langle
00\right| +e^{-i\gamma }\left| 01\right\rangle \left\langle 01\right|
+e^{i\gamma }\left| 02\right\rangle \left\langle 02\right|  \nonumber \\
&&+e^{i\gamma }\left| 10\right\rangle \left\langle 10\right| +e^{-i\gamma
}\left| 11\right\rangle \left\langle 11\right| +e^{i\gamma }\left|
12\right\rangle \left\langle 12\right|  \nonumber \\
&&+e^{-i\gamma }\left| 20\right\rangle \left\langle 20\right| +e^{i\gamma
}\left| 21\right\rangle \left\langle 21\right| +e^{-i\gamma }\left|
22\right\rangle \left\langle 22\right| ,
\end{eqnarray}
Afterward, we put $U_{A}=\left| 0_{A}\right\rangle \left\langle 0_{A}\right|
+\left| 1_{A}\right\rangle \left\langle 1_{A}\right| -\left|
2_{A}\right\rangle \left\langle 2_{A}\right| $, $U_{B}=\left|
0_{B}\right\rangle \left\langle 0_{B}\right| +\left| 1_{B}\right\rangle
\left\langle 1_{B}\right| -\left| 2_{B}\right\rangle \left\langle
2_{B}\right| $ and $\xi =\delta $, analogically we can get 
\begin{eqnarray}
U_{AB}^{\prime \prime }(\delta ) &=&e^{i\delta }\left| 00\right\rangle
\left\langle 00\right| +e^{i\delta }\left| 01\right\rangle \left\langle
01\right| +e^{-i\delta }\left| 02\right\rangle \left\langle 02\right| 
\nonumber \\
&&+e^{i\delta }\left| 10\right\rangle \left\langle 10\right| +e^{i\delta
}\left| 11\right\rangle \left\langle 11\right| +e^{-i\delta }\left|
12\right\rangle \left\langle 12\right|  \nonumber \\
&&+e^{-i\delta }\left| 20\right\rangle \left\langle 20\right| +e^{-i\delta
}\left| 21\right\rangle \left\langle 21\right| +e^{i\delta }\left|
22\right\rangle \left\langle 22\right| .
\end{eqnarray}
It is easily to check up that the non-local phase operation $P_{AB}$ can be
constructed as

\begin{equation}
P_{AB}=P_{A}P_{B}[U_{AB}^{\prime \prime }(\pi /6)S_{A}S_{B}U_{AB}^{\prime
\prime }(\pi /6)][U_{AB}^{^{\prime }}(\pi /3)S_{A}S_{B}U_{AB}^{^{\prime
}}(\pi /3)],
\end{equation}
where $P_{A}=\left| 0_{A}\right\rangle \left\langle 0_{A}\right| -\left|
1_{A}\right\rangle \left\langle 1_{A}\right| -\left| 2_{A}\right\rangle
\left\langle 2_{A}\right| $, $P_{B}=-\left| 0_{B}\right\rangle \left\langle
0_{B}\right| +\left| 1_{B}\right\rangle \left\langle 1_{B}\right| +\left|
2_{B}\right\rangle \left\langle 2_{B}\right| $ are the single qutrit phase
gates, and $S_{i}=\left| 0_{i}\right\rangle \left\langle 0_{i}\right|
+\left| 1_{i}\right\rangle \left\langle 2_{i}\right| +\left|
2_{i}\right\rangle \left\langle 1_{i}\right| $ with $i=A$, $B$ the single
qutrit swap gates between levels $\left| 1\right\rangle $ and $\left|
2\right\rangle $. The non-local operations $U_{AB}^{^{\prime }}(\pi /3)$ and 
$U_{AB}^{\prime \prime }(\pi /6)$ are given by eqs.(23) and (24) with $%
\gamma =\pi /3$, $\delta =\pi /6$ respectively.

At the end of this paper, we point out that, combining the technique in \cite
{Chen} with ours, we can easily generalize the result of eq.(9) to the case
of containing multiple systems and realize a non-local operation between
multiple spatially separated particles, each of them with arbitrary (maybe
different) dimensions. The concrete form for this non-local operation is

\begin{equation}
U_{A_{1}A_{2}\ldots A_{N}}(\xi )=\exp [i\xi U_{A_{1}}U_{A_{2}}\ldots
U_{A_{N}}],
\end{equation}
where $A_{i}$ with $i=1,2,\ldots ,N$ denote $N$ spatially separated
particles, and $U_{A_{i}}$ the corresponding local operators that satisfied $%
U_{A_{i}}^{+}=U_{A_{i}}$ and $%
U_{A_{i}}^{+}U_{A_{i}}=U_{A_{i}}U_{A_{i}}^{+}=1 $. The consumptions for
producing this multi-part and multi-dimensional non-local operation are $%
(N-1)$ pairs of classical bit communication and a non-local quantum
entangled qubit resource between $N$ partners, attached by some quantum
local unitary operations and quantum measurements. However, for the method
presented in reference \cite{Reznik}, there will be needed $N$ pairs of
entangled qudit resource along with $N$ pairs of classical dit communication.

In conclusion, we have proposed a method to implement a kind of non-local
operations between spatially separated two systems with arbitrary high
dimensions, by using only low-dimensional qubit quantum channels and
classical bit communications. The result may be generalized
straightforwardly to apply for multiple systems, each of them with arbitrary
dimensions. Compared with existed approaches, our method can economize many
classical resources when higher-dimensional and multiple systems involved,
and the needed low-dimensional quantum channels may be more easily
established in practice. We have also shown the construction of the
non-local quantum XOR gate for qutrit systems in terms of the obtained
non-local operations as well as some single qutrit local gates, which
suggests the universality of the obtained non-local operations in
multi-level distributed quantum computation and non-local quantum
information processing.

This work was supported by the National Fundamental Research Program Grant
No. 2001CB309310, the National Natural Science Foundation of China under
Grant Nos.10347128, 10325523 and 90203018, the Natural Science Foundation of
Hunan Province (04JJ3017), and the Scientific Research Fund of Hunan
Provincial Education Bureau (03C214, 03094, 02A026).


\begin{references}
\bibitem{Nielsen}  M. A. Nielsen and I. L. Chuang, {\it Quantum Computation
and Quantum Information}, (Cambridge University Press, Cambridge, England,
2004).

\bibitem{Eisert}  J. Eisert, K. Jacobs, P. Papadopoulos and M. B. Plenio,
Phys. Rev. A {\bf 62}, 052317 (2000).

\bibitem{Kraus}  B. Kraus and J. I. Cirac, Phys. Rev. A {\bf 63}, 062309
(2001).

\bibitem{Collins}  D. Collins, N. Linden and S. Popescu, Phys. Rev. A {\bf 64%
}, 032302 (2001).

\bibitem{Huelga}  S. F. Huelga, J. A. Vaccaro, A. Chefles and M. B. Plenio,
Phys. Rev. A {\bf 63}, 042303 (2001); S. Huelga, M. B. Plenio and J. A.
Vaccaro, Phys. Rev. A {\bf 65}, 042316 (2002).

\bibitem{Reznik}  B. Reznik, Y. Aharonov and B. Groisman, Phys. Rev. A {\bf %
65}, 032312 (2002).

\bibitem{Cirac}  J. I. Cirac, W. D\"{u}r, B. Kraus and M. Lewenstein, Phys.
Rev. Lett. {\bf 86}, 544 (2001); W. D\"{u}r, and J. I. Cirac, Phys. Rev. A 
{\bf 64}, 012317 (2001).

\bibitem{Dura}  W. D\"{u}r, G. Vidal, J. I. Cirac, N. Linden and S. Popescu,
Phys. Rev. Lett. {\bf 87}, 137901 (2001).

\bibitem{Durb}  W. D\"{u}r, G.Vidal and J. I. Cirac, Phys. Rev. Lett. {\bf 89%
}, 057901 (2002).

\bibitem{Huang}  Y. F. Huang, X. F. Ren, Y. S. Zhang, L. M. Duan and G. C.
Guo, Phys. Rev. Lett. {\bf 93}, 240501 (2004).

\bibitem{Berry}  B. Groisman and B. Reznik, Phys. Rev. A {\bf 71}, 032322
(2005).

\bibitem{Chen}  L. Chen and Y. X. Chen, quant-ph/0501107.

\bibitem{Cirac1}  J.I. Cirac, A.K.Ekert, S. F. Huelga and C. Macchiavello,
Phys. Rev. A {\bf 59}, 4249 (1999).

\bibitem{Stephen}  S. D. Bartlett, H. de Guise and B. C. Sanders, Phys. Rev.
A {\bf 65}, 052316 (2002).

\bibitem{Thew}  R. T. Thew, K. Nemoto, A. G. White and W. J. Munro, Phys.
Rev. A {\bf 66}, 012303 (2002).

\bibitem{Bruss}  D. Bruss and C. Macchiavello, Phys. Rev. Lett. {\bf 88},
127901 (2002); N. J. Cerf, M. Bourennane, A. Karlsson and N. Gisin, {\it ibid%
}. {\bf 88}, 127902 (2002).

\bibitem{Knill}  E. Knill, {\it Fault-tolerant postselected quantum
computation: schemes, }quant-ph/0402171 (2004).

\bibitem{Vedral}  V. Vedral, M. B. Plenio, M. A. Rippin and P. L. Knight,
Phys. Rev. Lett. {\bf 78}, 2275 (1997).

\bibitem{Klimov}  A. B. Klimov, R. Guzman, J. C. Retamal C. Saavedra, Phys.
Rev. A {\bf 67}, 062313 (2003).
\end{references}
\end{document}